\documentclass[twocolumn,prl, superscriptaddress,showpacs]{revtex4}
\usepackage{amsmath}
\usepackage{graphicx}
\usepackage{multirow}
\begin{document}

\title{Sub-atomic electronic feature from dynamic motion of Si dimer defects in bismuth nanolines on Si(001)}

\author{C. J. Kirkham}
\affiliation{National Institute for Materials Science (NIMS), Namiki 1-1, Tsukuba, Ibaraki, 305-0044, Japan}
\affiliation{London Centre for Nanotechnology and Department of Physics and Astronomy,University College London, London WC1E 6BT, United Kingdom}
\author{M. Longobardi}
\affiliation{Department of Quantum Matter Physics, University of Geneva, 24 Quai Ernest-Ansermet, CH-1211 Geneva 4, Switzerland}
\author{S. A. K\"{o}ster}
\affiliation{Department of Quantum Matter Physics, University of Geneva, 24 Quai Ernest-Ansermet, CH-1211 Geneva 4, Switzerland}
\author{Ch. Renner}
\affiliation{Department of Quantum Matter Physics, University of Geneva, 24 Quai Ernest-Ansermet, CH-1211 Geneva 4, Switzerland}
\author{D. R. Bowler}
\affiliation{London Centre for Nanotechnology and Department of Physics and Astronomy,University College London, London WC1E 6BT, United Kingdom}
\affiliation{International Center for Materials Nanoarchitectonics (MANA), National Institute for Materials Science (NIMS), Namiki 1-1, Tsukuba, Ibaraki, 305-0044, Japan}


\begin{abstract}
  Scanning tunneling microscopy (STM) reveals unusual sharp features in otherwise defect free bismuth nanolines self-assembled on Si(001). They appear as subatomic thin lines perpendicular to the bismuth nanoline at positive biases and as atomic size beads at negative biases. Density functional theory (DFT) simulations show that these features can be attributed to buckled Si dimers substituting for Bi dimers in the nanoline, where the sharp feature is the counterintuitive signature of these dimers flipping during scanning. The perfect correspondence between the STM data and the DFT simulation demonstrated in this study highlights the detailed understanding we have of the complex Bi-Si(001) Haiku system.
\end{abstract}

\maketitle
\begin{figure}[ht!]
\centering  
\includegraphics[width=\columnwidth]{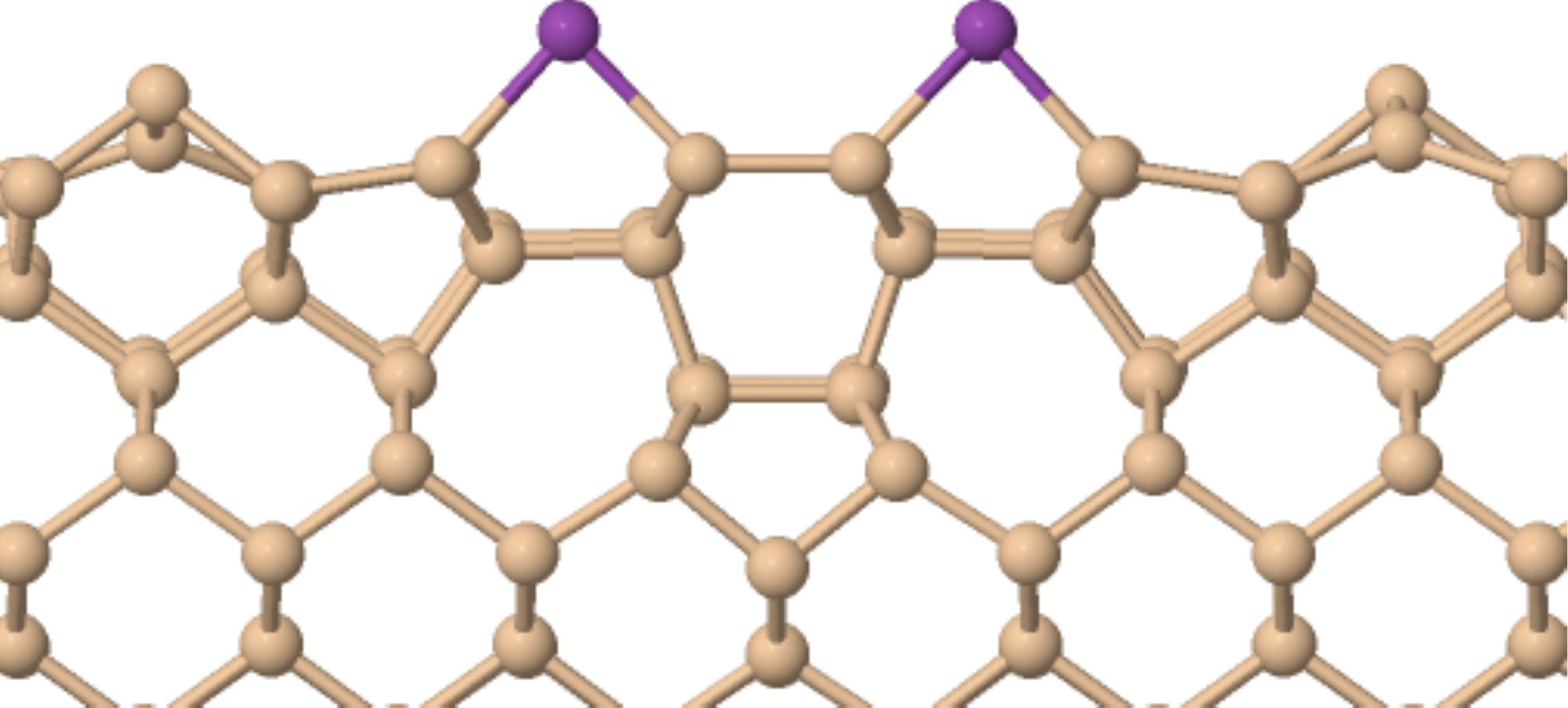} 
\caption{(Color online) Cross-section of the Haiku model for the Bi nanoline on Si(001). It consists of a reconstructed Si region containing a 5-7-5 membered ring arrangement, and two rows of Bi dimers. Si atoms are beige (light) and Bi atoms purple (dark).} 
\label{fig:Haiku} 
\end{figure}

Silicon is an important material in the semiconductor industry, and has recently attracted interest for nanoscale electronics. This includes single atom transistors~\cite{Fuechsle2012}, self-assembled Bi nanolines on Si(001)~\cite{Miki1999feb,Bowler2000,Owen2006}, and patterning of Si Dangling Bonds (DBs) on H:Si(001)~\cite{Kawai2012,Schofield2013,Bianco2013,Kolmer2014,Kleshchonok2015}. Bi nanolines are self-assembled atomically straight structures which can grow up to several micrometers long, and run perpendicular to the Si dimer rows. They consist of two rows of Bi dimers, which sit atop a reconstructed region of Si known as the Haiku structure, due to its 5-7-5 membered ring arrangement (Fig~\ref{fig:Haiku}). Elsewhere we have fully characterized the scanning tunneling microscopy (STM) appearance of the nanoline, and how it varies based on the arrangement of the surrounding Si surface~\cite{STM}. It has been shown that a H cracker can be used to remove the Bi without disrupting the underlying Haiku structure~\cite{Bianco2011}, giving a long straight template for assembling other atomic species, known as the Haiku stripe.

The patterning of Si DBs on H:Si(001) allows for the study of systems not accessible on the clean surface, such as isolated DBs and clean dimers. A symmetric butterfly-like appearance has been observed in empty state STM images of isolated dimers, despite being physically asymmetrically buckled~\cite{Chen2002,Bellec2009}. The butterfly-like appearance consists of two dominant round outer lobes, and a central sharp lobe oriented perpendicular to the Si dimer. Similar features are observed for Ge dimers on H:Ge(001), but with a more distinct central lobe. Such butterflies have not been observed on clean surfaces. Recently these features were successfully reproduced in simulations~\cite{Engelund2016}, and explained in terms of rapid tip-induced flipping between equivalent dimer buckling configurations, through an unoccupied DB dimer gap state. Above a certain bias, the flipping rate is faster than the time resolution of the STM, giving rise to the symmetric appearance. The central sharp lobe is due to the physical and electronic asymmetry of the two buckling configurations. The weaker central lobe observed on Si compared to Ge is a consequence of a weaker DB asymmetry in Si.

Here we present a new system which combines isolated Si dimers and Bi nanolines. Si dimers appear as unusual defects on otherwise perfect Bi nanolines, which STM  reveals as a sharp line at positive bias and as a round spot at negative bias. We show that this distinct footprint can be understood as a buckled Si dimer that flips during scanning. Compared with Si dimers on H:Si(001), the central sharp feature is significantly more distinct, suggesting an effect related to the Haiku substrate. This system offers a chance to study a unique 1D electronic system.

The experiments were carried out in an ultrahigh vacuum (UHV) LT-STM by Omicron (base pressure less than 2$\times$10$^{-11}$~mBar), equipped with in situ facilities for sample heating and Bi evaporation. The samples were cut from commercial polished p-type Si(001) wafers (Boron doped, 0.1~$\Omega$cm). Before insertion into UHV, the silicon was degreased in ultrasonic acetone followed by isopropanol baths (5 minutes each) and rinsed in purified water.  The silicon surface was then immersed in an oxidation mixture of H$_2$O$_2$ and H$_2$SO$_4$ (1:1 by volume) for 30 seconds, followed by a HF (15 wt\%) dip for 30 seconds. This process was repeated many times to grow a thin protective oxide film onto the silicon surface.

In the STM preparation chamber, the Si(001) was re-cleaned by direct current outgassing at 700$^{\circ}$C for 12~h followed by many cycles of flashing up to 1200$^{\circ}$C. The base pressure of the UHV chamber was kept below 2$\times$10$^{-9}$~mBar during both the outgassing and flashing. The quality of the Si(001) surface was controlled by monitoring the reflection high-energy electron diffraction (RHEED) pattern in real time during the heating process. Bismuth was evaporated from a ceramic crucible (MBE Komponenten K-cell heated at 470$^{\circ}$C) onto the clean silicon substrate kept at 450$^{\circ}$C for 12 minutes followed by annealing at the same temperature for 3 minutes. The formation of the Bi nanolines was monitored by the characteristic arc connecting the Si spots in the RHEED pattern, usually appearing 5 to 8 minutes into  the deposition process. The Si(001) sample with nanolines was then transferred into the STM and all the topographs were taken in the constant current mode at 77~K. The bias voltage was applied to the sample and we used cut PtIr wire tips. 

DFT as implemented in the Vienna \textit{Ab initio} Simulation Package (VASP)~\cite{Kresse1996} version 5.4.1 was used to perform the calculations. Ultrasoft pseudopotentials (US-PP)~\cite{Vanderbilt1990} with the PW91 exchange-correlation functional~\cite{Perdew1992} were used. For Bi the $6s^26p^3$ electrons were treated as valence, the rest as core.

\begin{figure}[ht!]
\centering  
\includegraphics[width=\columnwidth]{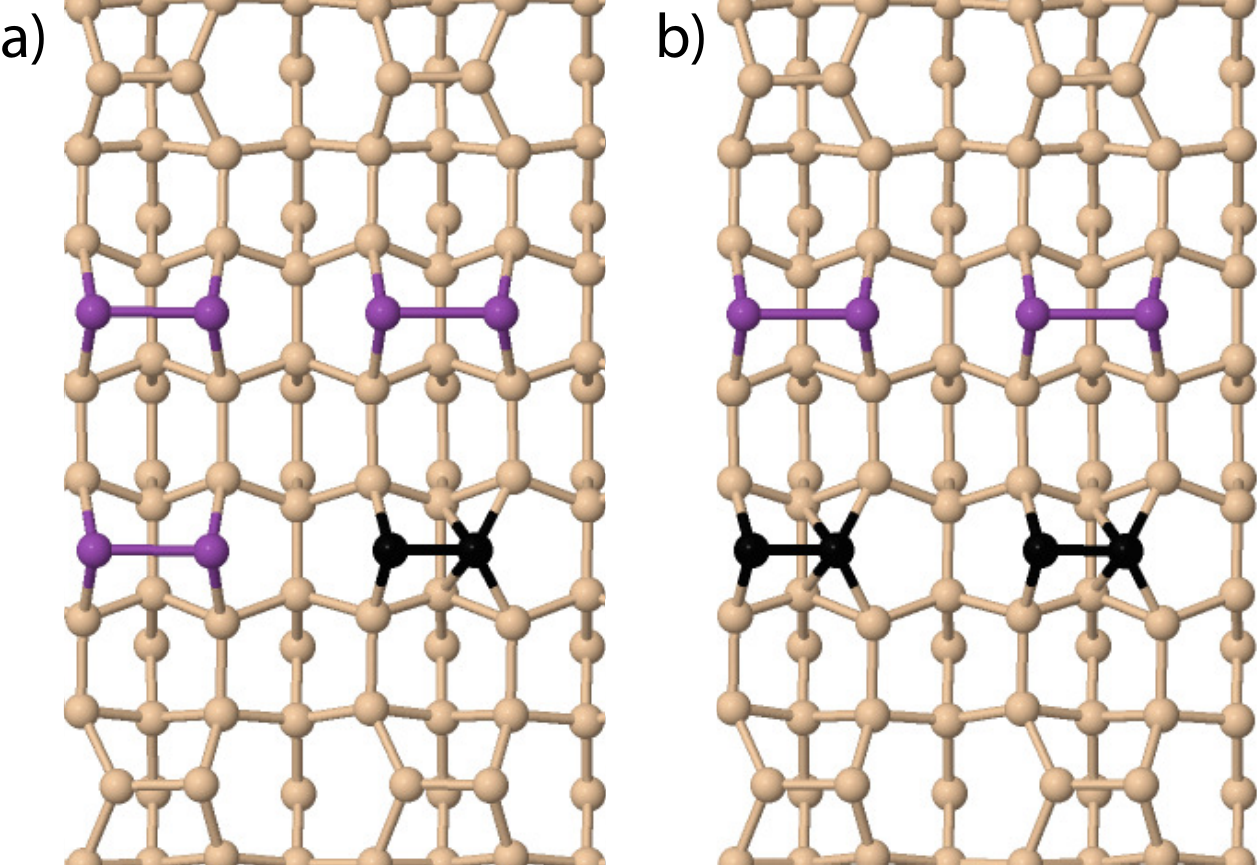} 
\caption{(Color online) Defective Bi nanolines containing a) one isolated and b) two adjacent buckled Si dimers. For clarity only the first four Si layers are shown. Si atoms are beige (light), Bi atoms purple (dark) and Si dimer defects black.} 
\label{fig:Theory} 
\end{figure}

The Si(001) surface was represented by a periodically repeated 10 layer slab model with a reconstructed surface layer consisting of buckled Si dimers with p(2$\times$2) periodicity. The Si surface consisted of two rows of Si dimers, made up of six regular dimers and the Haiku reconstructed region. The Bi nanoline consists of a pair of Bi dimers per dimer row sitting atop the Haiku core. For defective nanolines, we replaced one or two of the Bi dimers along the nanoline direction with buckled Si dimers, as shown in Fig.~\ref{fig:Theory}. These dimers either maintained the surface buckling pattern, or broke it, which we termed Left and Right buckling respectively.

The experimental bulk lattice constant for Si of $a_o=5.4306$~\AA~was used. The bottom Si layer was terminated by H atoms in a dihydride structure. Both the H atoms and the bottom two layers of Si atoms were fixed, in order to simulate a bulk like environment, with all other atoms allowed to move freely. The periodic images of the surface were separated by a vacuum gap of 12.73~\AA~to prevent interaction between repeated images. An energy cut-off of 250 eV was used. A (2$\times$1$\times$1) Monkhorst-Pack k-mesh was used with the largest number of k-points along the direction of the Bi nanoline.  Geometry optimizations were performed with a 0.02~eV/\AA~convergence condition for the forces on each atom. All calculations were spin polarized with no restrictions placed on the spins.

To compare the viability of defective nanolines against perfect nanolines we calculated
 
\begin{equation} 
\label{eq:Ediff} 
E_{V}=E_{\text{Defect}}-2\times E_{Si} + E_{Bi}, 
\end{equation} 
where $E_{\text{Defect}}$ is the energy of the defective nanoline containing Si, $E_{Si}$ is the energy of a Si atom, calculated as $\frac{1}{512}$ the energy of a 512 atom bulk cell, and $E_{Bi}$ is the energy of a Bi dimer reservoir, calculated as the energy of a Bi dimer on the clean p(2$\times$2) surface compared against a bare surface. E$_V$ can be compared directly against the energy of the perfect nanoline.

Simulated STM images were produced using the Tersoff-Hamman method, as implemented in bSKAN33~\cite{Hofer2003}. We applied the techniques developed by Engelund $et~al.$~\cite{Engelund2016} for Si dimers on H:Si(001) and combined the currents as follows
 
\begin{equation} 
  \label{eq:Current} 
\langle I(r) \rangle = \frac{I_{DB,R}(r)I_L(r)+I_{DB,L}(r)I_R(r)}{I_{DB,L}(r)+I_{DB,R}(r)}, 
\end{equation} 
where L/R refer to the Left and Right buckling configurations of the Si dimers, and DB refers to the dangling bond states. To isolate the DB states from the full spectrum and find a single state with majority contributions from the DB, we examined the band decomposed charge densities around the Fermi Energy ($E_F$).


\begin{figure}[ht!]
\centering  
\includegraphics[width=\columnwidth]{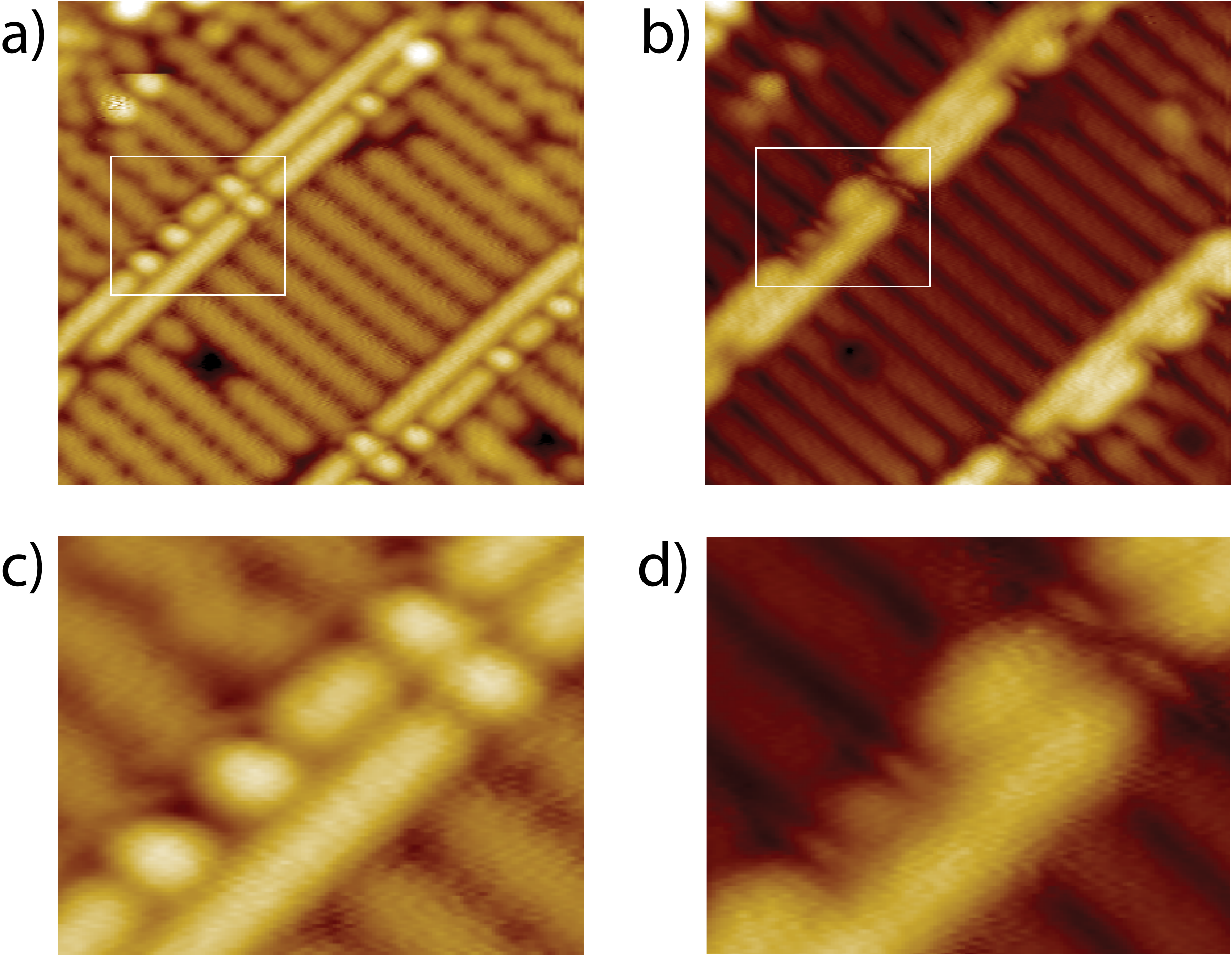} 
\caption{(Color online) 8.35$\times$7.45~nm$^2$ STM images of self-assembled Bi nanolines with defects on Si(001) measured at a) -3.0~V and b) +1.8~V sample bias. c) and d) 2.69$\times$2.13~nm$^2$ magnified images of the defects highlighted by the rectangular boxes.} 
\label{fig:Expt} 
\end{figure}

Figure~\ref{fig:Expt} shows high resolution STM micrographs of two Bi nanolines measured at positive and negative bias. Of particular interest are the striking defects emphasized in Figs.~\ref{fig:Expt} c) and d), which are either isolated or cluster along the Bi dimer rows with a nearest distance corresponding to the spacing between surface Si dimers. They can also be found on adjacent Bi dimer rows, suggesting they nucleate independently on each parallel dimer row of the Bi nanoline.    

At negative bias, the defects appear as beads with the same brightness as the host Bi nanoline. They are aligned on the Bi dimer rows and centered on the dimers of the Si(001) surface. Each bead is flanked by two dark regions that are most prominent between adjacent defects on the same Bi dimer row. These dark regions are reminiscent of the dark trenches separating the Si dimer rows on the reconstructed Si(001) surface.  

The defects are most striking at positive bias, where they appear as very sharp, nearly subatomic, features running perpendicular to the Bi dimers. They are darker than the Bi dimer rows and become brighter as the bias is reduced. The defects are centered on the Si dimer rows of the Si(001) surface, although the imaging contrast in Figs.~\ref{fig:Expt} b) and d) suggests they are aligned with the trenches between Si dimers. This false impression is due to a known phase shift of the imaging contrast which makes the center of the dimer rows appear dark at this particular bias~\cite{Hata1999}. Adjacent narrow features are separated by an additional rounder feature which aligns with the trench (Fig.~\ref{fig:Expt} d)).

Most of the time, Bi nanolines are defect free and the nature of these features was a real puzzle. We initially tried to understand them in terms of an adsorbed contaminant, but all adsorbates failed to reproduce the experimental results - full details are given in the Supplemental Material. The puzzle was solved when we realized that the topographic signature of the defects at positive bias is very similar to that of isolated Si dimers on H:Si(001). Indeed, distinct perpendicular lines are observed in the middle of these dimers, although less pronounced than on the Bi nanoline defects, prompting us to consider embedded Si dimers to explain the defects in the Bi nanolines.

Guided by the similar bias dependence of the Bi nanoline defects and Si dimers on H:Si(001)~\cite{Engelund2016}, we examined the case of a single Si dimer substituted for one of the Bi dimers as shown in Fig.~\ref{fig:Theory} a). This could occur, for example, if a Si dimer were to fill a missing dimer defect in the nanoline. The Left buckled configuration was found to be 0.06~eV lower in energy. This configuration is lower in energy because it continues the surface buckling pattern, and the difference between a Left and Right buckled Si dimer is small enough for STM tip induced flipping to occur. 

To assess the viability of these defective nanolines, we compared E$_{V}$ as specified in Eq.~\ref{eq:Ediff} against the energy of the perfect nanoline. For the Left buckled configuration, the perfect nanoline was 0.6~eV lower in energy. This means that whilst filling vacancies in the nanoline with Bi dimers is still favored, it would not be unreasonable for Si dimers to fill it instead.

\begin{figure}[ht!]
  \centering
      \includegraphics[width=\columnwidth]{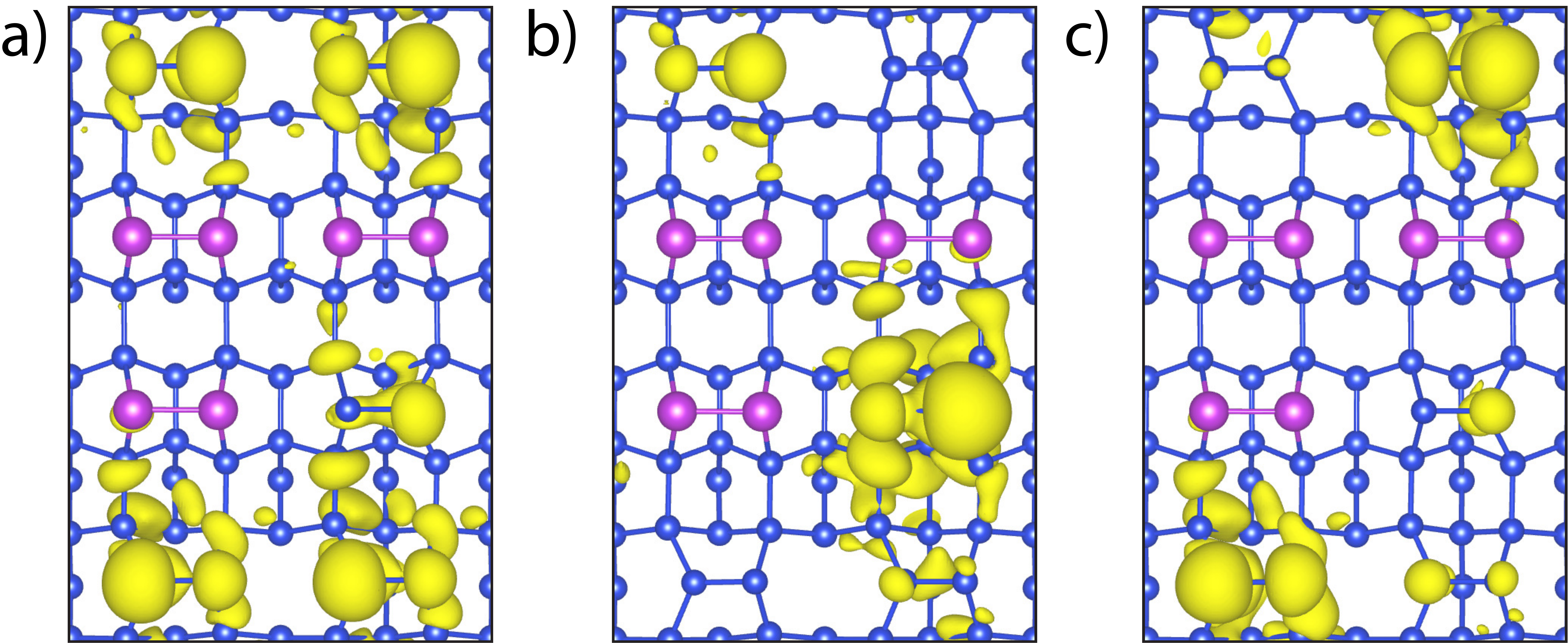} 
      \caption{(Color online) Band decomposed charge densities for a defective Bi nanoline containing a single Left buckled Si dimer. The a) $5^{th}$, b) $6^{th}$ and c) $7^{th}$ unoccupied bands above $E_F$ are shown. Si atoms are blue, Bi atoms purple, and charge density is depicted in yellow.} 
\label{fig:Bands} 
\end{figure}
 
With the sustainability of Si defects in the nanolines established, we examined the band decomposed charge densities for the unoccupied bands close to $E_F$, some of which are shown in Fig.~\ref{fig:Bands} for the Left buckled case. In nearly all cases, the most significant contributions come from the surface Si dimers, except for the sixth unoccupied band, at 0.74~eV above $E_F$, where the most significant contribution is from the Si dimer within the nanoline. We assigned this state, and the equivalent Right buckled state, to $I_{DB}$, in order to calculate the modified currents specified in Eq.~\ref{eq:Current}. We applied the same methods for the occupied bands, and to a Bi nanoline containing two adjacent Si dimer defects on the same Bi dimer row.

\begin{figure}[ht!]
\centering   
\includegraphics[width=\columnwidth]{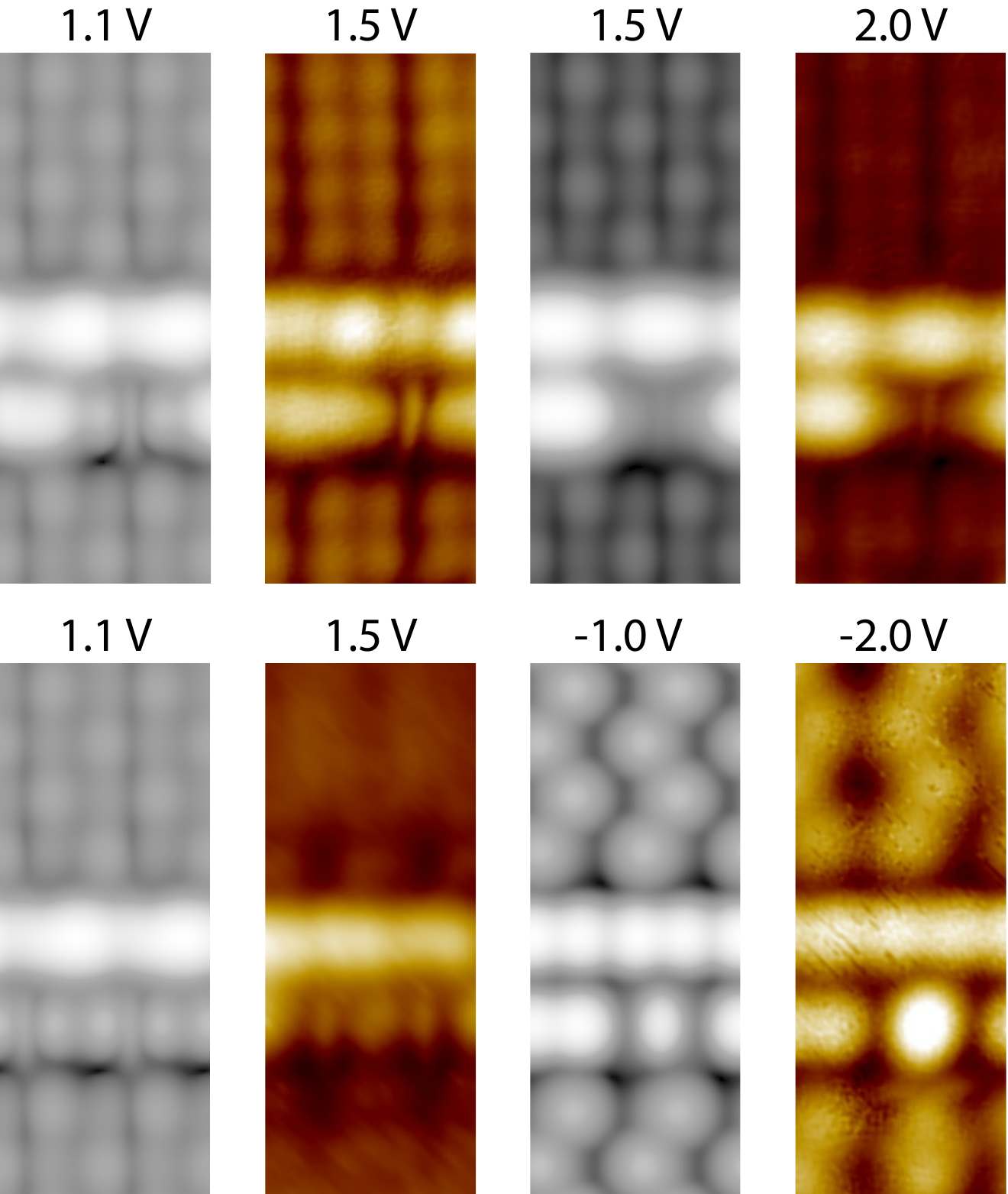} 
\caption{Simulated (gray scale) and corresponding actual 1.54$\times$3.84~nm$^2$ STM images (color scale) of a Bi nanoline containing a single or two adjacent buckled Si dimer defects. The simulated images are based on the modified currents from Eq.~\ref{eq:Current}. The STM images set points are (1.5~V; 2~nA), (2.0~V; 1~nA), (1.5~V; 0.2~nA), and (-2.0~V; 0.1~nA).}
\label{fig:STM} 
\end{figure}

The resulting simulated STM images are shown in Fig.~\ref{fig:STM}. A distinct sharp feature is observed down the center of the Si dimer over a range of positive biases. It is more visible than the equivalent feature in the H:Si(001) case~\cite{Engelund2016}. The Si background phase shift is reproduced above 1.0~V with a dark line appearing down the center of the Si dimers, aligning with the sharp feature. As the bias is increased further the contrast of the Si dimer feature decreases compared to the surrounding Bi dimers and becomes barely visible above 1.5~V. For comparisons to experiment it should be noted that the biases do not match exactly, with the equivalent simulated results appearing about 0.5~V lower. With this in mind, there is excellent agreement between simulations and experiment. At negatives bias we observed a round feature centered on the dimer row, again in excellent agreement with experiment.

The general behavior for two adjacent Si dimer defects on the same Bi dimer row is the same as for a single dimer, with sharp lines down the Si dimer centers, fading out with increased bias. An additional rounder feature is observed between each pair of narrow lines, which looks like a cross-trench dimer, but is actually Si atoms from two neighboring dimers. The same type of phase shift occurs for the Si background. There is good agreement with experiment, where a rounder feature was observed between adjacent sharp features that is aligned with the trench. These simulations are equivalent to replacing half the nanoline with Si, suggesting there is no size limit to this behavior. Experimentally, we observed up to four adjacent Si dimer defects on a single Bi dimer row.

\begin{figure}[ht!]
\centering   
\includegraphics[width=\columnwidth]{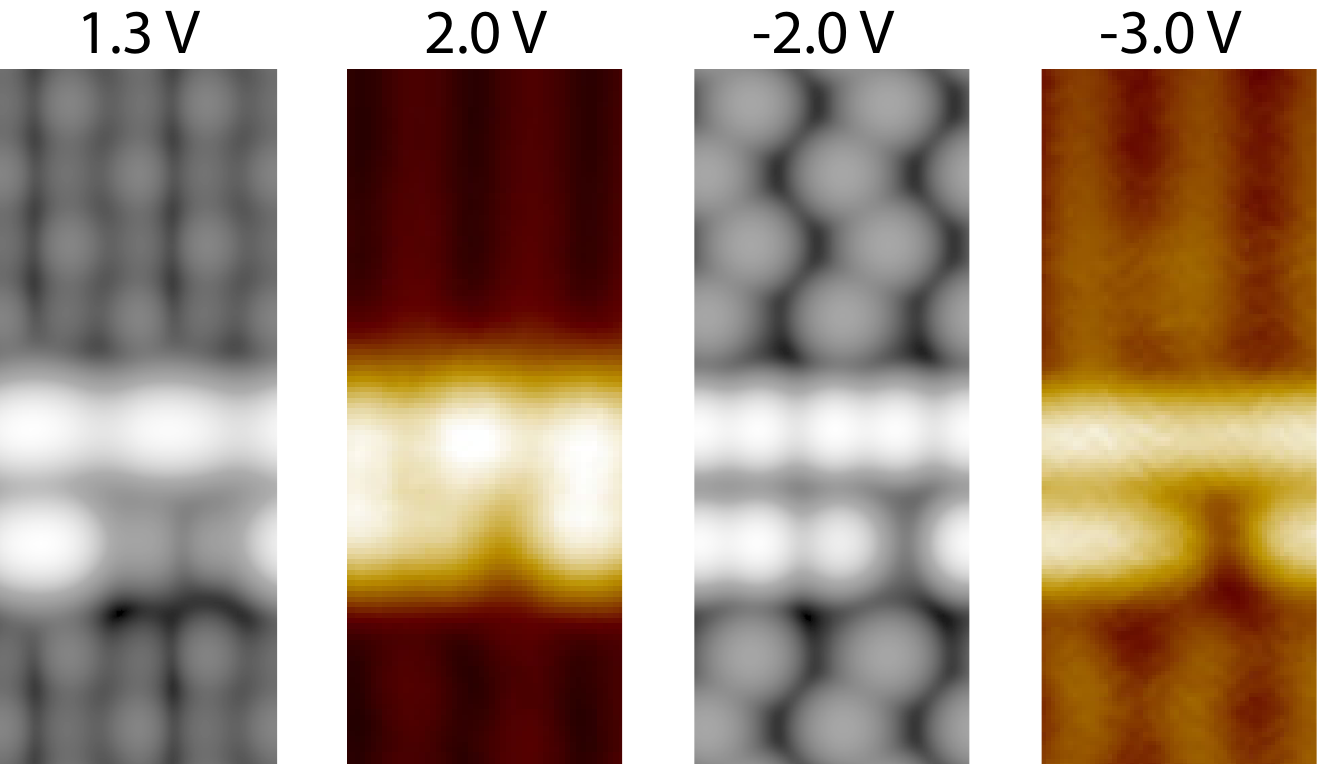} 
\caption{
Simulated (gray scale) and corresponding actual 1.54$\times$3.84~nm$^2$ STM images (color scale) of a Bi nanoline containing a single buckled non-flipping Si dimer defect. The simulation is based on non-modified currents corresponding to a frozen dimer defect. The STM images set points are (2.0~V; 0.1~nA), and (-3.0~V; 0.2~nA).}
\label{fig:Frozen_STM} 
\end{figure}

The majority of defects correspond to the flipping dimers shown in Fig.~\ref{fig:STM}. On rare occasions we see defects with a different contrast, consisting of an asymmetric bright and dark dimer structure at both polarities (Fig.~\ref{fig:Frozen_STM}). This defect is perfectly reproduced using the same models, but with Si dimer defects which do not flip during scanning. We have observed both defect types in the same STM images, but the reason for the rare instances of frozen dimers is not clear. A possibility are nearby defects and impurity atoms in the surrounding Si.  

Having confirmed the identity of the defects, simulations can explain why the central feature is more distinct here than for H:Si(001). It was initially believed that confinement between Bi dimers might contribute, but the appearance of the sharp feature does not change even when half of the nanoline is replaced with Si dimers in the model. Instead it is possible to compare the physical and electronic asymmetry against surface dimers, because as noted earlier, the greater the dimer asymmetry, the more distinct the central feature. We found that the buckling of the Si dimers is indeed slightly more pronounced on the nanoline than on the Si surface, with an increase of 7\% in vertical separation, and 5\% in buckling angle. Additionally, their electronic asymmetry is slightly larger as shown in Fig.~\ref{fig:Bands}.

Combining DFT modeling and high resolution STM imaging, we identify a new Si dangling bond system embedded in self-assembled Bi nanolines on the bare Si(001) surface. We analyze unusual sharp defects imaged by STM on the nanolines. Their generic shapes at positive and negative bias bear striking similarities with Si dimers found on H:Si(001). However, in contrast to the H:Si(001) case, the narrow feature observed here on the hydrogen free Haiku structure is significantly sharper. Using DFT modeling, we identify these defects as Si dimers embedded in the Bi nanoline. We show that the narrow feature is a direct consequence of the defect dimer flipping during STM scanning. Indeed, we observe a few rare instances of asymmetric defects without the narrow feature that can be simulated assuming a frozen dimer in the same model description. This confirms the very counterintuitive conclusion that the subatomic sharp feature is a direct consequence of imaging a dynamic structure during the STM scan~\cite{Engelund2016}. The perfect correspondence between the STM data and the DFT simulation demonstrated in this study highlights the detailed understanding we have of the complex Bi-Si(001) Haiku system.

We thank F. Bianco, H. Zandvliet and J. Owen for stimulating discussions. We are grateful to T. Miyazaki for computer time to simulate the STM images. We thank G. Manfrini for his technical assistance in the STM laboratories. D.R.B. and C.J.K. acknowledge financial support from a UCL Impact Studentship; C.R., M.L. and S.A.K. acknowledge financial support from the Swiss National Science Foundation.

\bibliography{Feature}
\end{document}